\documentclass[aps,prc,reprint,groupedaddress,amsmath,amssymb,floatfix]{revtex4-2}

\usepackage[T1]{fontenc}
\usepackage{graphicx}
\usepackage{bm}
\usepackage{hyperref}
\hypersetup{colorlinks=true,citecolor=blue,linkcolor=blue,urlcolor=blue,pdftitle={Relative geometry of nuclear mean fields through differential forms},pdfauthor={Rafael C. R. de Lima}}

\newcommand{\Nsol}{98}
\newcommand{\symOK}{all}
\newcommand{\valSrad}{2.4\times10^{-20}}
\newcommand{\valSdep}{6.7\times10^{-8}}
\newcommand{\valErrA}{2.0\times10^{-2}}
\newcommand{\valErrB}{4.9\times10^{-3}}
\newcommand{\valErrC}{1.2\times10^{-3}}
\newcommand{\valRatio}{3.99}
\newcommand{\valRich}{3.1\times10^{-6}}
\newcommand{\TFnull}{1.1\times10^{-16}}
\newcommand{\corrSP}{0.93}
\newcommand{\invSeries}{0.999}
\newcommand{\invSeriesSd}{0.010}
\newcommand{\Spnseven}{1.5\times10^{-4}}
\newcommand{\Spnthree}{3.9\times10^{-5}}
\newcommand{\qseven}{0.023}
\newcommand{\qthree}{0.014}
\newcommand{\Ssw}{5.5\times10^{-7}}
\newcommand{\Snsig}{3.6\times10^{-6}}
\newcommand{\errFFT}{3.3\times10^{-10}}
\newcommand{\errFD}{23}
\newcommand{\SratioMg}{253}
\newcommand{\SmgThreeFour}{4.5\times10^{-3}}

\newcommand{\skinMg}{0.39}
\newcommand{\nnAngZero}{13}
\newcommand{\nnAngSeven}{1.3}
\newcommand{\npAngZero}{18}
\newcommand{\npAngSeven}{2.2}
\newcommand{\tnMg}{1.13}
\newcommand{\tpMg}{0.73}
\newcommand{\tNZmin}{0.80}
\newcommand{\tNZmax}{0.88}
\newcommand{\drThreeMg}{+0.18}
\newcommand{\drSevenMg}{-0.22}
\newcommand{\drNZmax}{0.01}
\newcommand{\anMgMin}{0.05}
\newcommand{\anMgMax}{0.07}
\newcommand{\anNZmax}{0.02}
\newcommand{\logRatioMin}{115}
\newcommand{\logRatioMax}{358}
\newcommand{\logVar}{8}
\newcommand{\naxAxMin}{1.1\times10^{-3}}
\newcommand{\naxAxMax}{2.7\times10^{-3}}
\newcommand{\naxTriMin}{2.1\times10^{-2}}
\newcommand{\naxTriMax}{2.4\times10^{-2}}

\newcommand{\dtTriMin}{0.15}

\newcommand{\SpnGe}{8.3\times10^{-4}}
\newcommand{\SpnXe}{3.4\times10^{-4}}
\newcommand{\fermiOne}{2.5\times10^{-4}}
\newcommand{\fermiEight}{3.0\times10^{-5}}
\newcommand{\fermiSix}{1.4\times10^{-6}}
\newcommand{\fermiHalf}{6.9\times10^{-7}}
\newcommand{\coulFFTc}{5.5\times10^{-4}}
\newcommand{\coulFFTthree}{2}
\newcommand{\coulFFTsix}{8}
\newcommand{\coulFDtwo}{4}
\newcommand{\coulFDfour}{0.5}
\newcommand{\coulFDfourHalf}{0.04}
\newcommand{\oriAnaOne}{4.2\times10^{-7}}
\newcommand{\oriAnaHalf}{7.2\times10^{-12}}
\newcommand{\oriTriple}{0.024}
\newcommand{\oriMgS}{1.0}

\newcommand{\oriMgB}{0.049}
\newcommand{\collEcm}{40}
\newcommand{\collB}{2.5}
\newcommand{\collT}{300}
\newcommand{\collDt}{0.2}
\newcommand{\collGrid}{48\times32\times48}
\newcommand{\collTpre}{90}
\newcommand{\collSpnPre}{2.5\times10^{-3}}
\newcommand{\collSpnPreVar}{3}
\newcommand{\collTWstart}{1.3\times10^{-3}}
\newcommand{\collTWpre}{1.2\times10^{-2}}
\newcommand{\collSwVar}{3}

\newcommand{\collWstart}{3.6\times10^{-6}}
\newcommand{\collWpre}{3.1\times10^{-5}}
\newcommand{\collSeppre}{8}
\newcommand{\collWmax}{6.9\times10^{-5}}
\newcommand{\collWpostMin}{2.0\times10^{-5}}
\newcommand{\collWpostMax}{4.6\times10^{-5}}
\newcommand{\collTWpostMin}{4.2\times10^{-3}}
\newcommand{\collTWpostMax}{1.3\times10^{-2}}

\newcommand{\collSpnPostMin}{2.1\times10^{-3}}
\newcommand{\collSpnPostMax}{5.4\times10^{-3}}
\newcommand{\SnpWMg}{1.8\times10^{-7}}
\newcommand{\cvSpn}{1.3}
\newcommand{\cvSpnBox}{0.03}
\newcommand{\cvSpW}{5}
\newcommand{\cvE}{0.03}
\newcommand{\cvBeta}{0.003}
\newcommand{\cvRatio}{253}
\newcommand{\cvNaxAxMin}{1.8\times10^{-5}}
\newcommand{\cvNaxAxMax}{5.6\times10^{-5}}
\newcommand{\cvNaxGe}{2.4\times10^{-2}}

\newcommand{\cvWGeVar}{5}

\newcommand{\cvWSep}{1300}

\newcommand{\thickFine}{0.01}

\newcommand{\tddGap}{400}
\newcommand{\collConnLow}{90}
\newcommand{\collConnHigh}{110}
\newcommand{\invSeriesMax}{0.057}
\newcommand{\lvPeakMgTwo}{0.66}
\newcommand{\lvPeakMgThree}{0.51}
\newcommand{\lvPeakGe}{0.49}
\newcommand{\lvRatioMin}{77}
\newcommand{\lvRatioMax}{430}
\newcommand{\lvDev}{3}
\newcommand{\cosMg}{3.7\times10^{-4}}
\newcommand{\cosGe}{0.034}
\newcommand{\cosGeMax}{0.12}
\newcommand{\TWaxMin}{4.0\times10^{-5}}
\newcommand{\TWaxMax}{1.6\times10^{-4}}
\newcommand{\TWtriMin}{9.4\times10^{-4}}
\newcommand{\TWtriMax}{2.6\times10^{-3}}
\newcommand{\TWaxSixMin}{1.2\times10^{-7}}
\newcommand{\TWaxSixMax}{3.7\times10^{-7}}
\newcommand{\TWGeOne}{9.4\times10^{-4}}
\newcommand{\TWGeSix}{8.9\times10^{-4}}
\newcommand{\TWsepOne}{6}
\newcommand{\TWsepSix}{2400}
\newcommand{\SpwMgThree}{4.5\times10^{-3}}
\newcommand{\coareaOne}{0.9}
\newcommand{\coareaSix}{0.4}
\newcommand{\Tfloor}{6.8\times10^{-14}}
\newcommand{\lvSigPeak}{0.025}
\newcommand{\lvSigS}{21}
\newcommand{\lvSigSeight}{6}
\newcommand{\lvSigInt}{1.0}
\newcommand{\lvUpS}{0.7}
\newcommand{\lvRatioVarMin}{72}
\newcommand{\lvRatioVarMax}{440}
\newcommand{\chkBasisSeven}{3.2}
\newcommand{\chkBasisTen}{0.3}
\newcommand{\chkMeshSeven}{7.3}
\newcommand{\chkPairsMesh}{12}
\newcommand{\chkPairsBasis}{3.5}
\newcommand{\chkEA}{1.3}
\newcommand{\chkQp}{4.5}
\newcommand{\startLow}{13}
\newcommand{\startLowMax}{17}
\newcommand{\startHigh}{4.3}
\newcommand{\startTen}{3.7}
\newcommand{\startZeroOb}{6.0\times10^{-5}}
\newcommand{\startZeroPr}{1.2\times10^{-4}}
\newcommand{\qZeroOb}{-0.21}
\newcommand{\qZeroPr}{+0.24}
\newcommand{\dEZero}{10}
\newcommand{\solNF}{20}
\newcommand{\solKmax}{3.5}
\newcommand{\solKT}{0.2}
\newcommand{\solMix}{0.3}
\newcommand{\solMesh}{80\times28}
\newcommand{\qTol}{0.02}
\newcommand{\qContrast}{5.9}
\newcommand{\qfourthree}{-0.016}
\newcommand{\qfourseven}{+0.006}
\newcommand{\collTdisc}{85}
\newcommand{\collWfacDisc}{7.3}
\newcommand{\collTWfacDisc}{7.5}
\newcommand{\collWfacExact}{8.6}
\newcommand{\collTWfacExact}{8.8}
\newcommand{\geLongIter}{15000}
\newcommand{\geConvIter}{13100}
\newcommand{\geGammaMin}{8.5}
\newcommand{\geGammaMax}{8.7}
\newcommand{\geLongDrift}{2.7\times10^{-7}}
\newcommand{\poissonCheck}{0.7}
\newcommand{\rigSep}{8.2}
\newcommand{\rigTWfac}{11.5}
\newcommand{\dynTWfacP}{8.9}

\newcommand{\rigSwVar}{15}
\newcommand{\convDtMax}{1.6}
\newcommand{\convDxMax}{5.5}
\newcommand{\convSepMax}{0.33}
\newcommand{\convGridFine}{60\times40\times60}
\newcommand{\convWfacDiscFine}{7.7}
\newcommand{\convTWfacDiscFine}{7.6}
\newcommand{\convSwVarFine}{1}
\newcommand{\corrMagMin}{79}
\newcommand{\corrMagMax}{93}

\newcommand{\corrGeXe}{2.4}
\newcommand{\corrCellMag}{98}

\newcommand{\corrCellPairs}{4.9}

\newcommand{\nearBzero}{2.0}

\newcommand{\nearPairs}{31}
\newcommand{\nearMax}{1.21}
\newcommand{\nearSpread}{17}

\begin{document}
\raggedbottom

\title{Relative geometry of nuclear mean fields through differential forms}

\author{Rafael C. R. de Lima}
\email[Contact author: ]{rafael.nuclear@gmail.com}
\affiliation{Universidade do Estado de Santa Catarina (UDESC), Santa Catarina, Brazil}

\date{\today}

\begin{abstract}
\begin{description}

\item[Background] Proton and neutron densities can have similar multipole moments while differing locally. Comparing them requires a measure of how their level surfaces meet, including in periodic nuclear structures with different spatial connectivity.

\item[Purpose] We use differential forms to measure and locate differences between nuclear mean fields, and to distinguish the misalignment of two density gradients from the variation of a third field along their level-surface intersections.

\item[Method] A pair invariant measures the angle between two gradients, weighted by their magnitudes. The coarea formula resolves its contribution over level surfaces. A triple invariant factorizes exactly into a weighted pair term and a directional term $T_W$ for the third field. We test these quantities in \Nsol{} magnetized Wigner--Seitz solutions, five static Hartree--Fock configurations, an illustrative $^{16}$O\,+\,$^{34}$Mg collision, and periodic nuclear pasta configurations comprising a slab and a connected network.

\item[Results] The pair invariant distinguishes cell solutions with comparable proton quadrupole moments. It is \SratioMg{} times larger in $^{34}$Mg than in $^{24}$Mg, with the difference distributed over the proton surfaces; refinement to $0.6$ fm changes it by at most $\cvSpn\%$ in the four configurations recomputed. The directional factor separates the sampled triaxial configurations from the prolate ones. In the collision, it accounts for the $\collWfacDisc$-fold rise of the triple before the first recorded low-density connection, while the weighted pair term changes by at most $\collSwVar\%$. Rigidly approaching static fragments produce a comparable rise. In periodic nuclear pasta, slab and connected-network configurations exhibit similar integrated proton--neutron misalignment but different distributions over density levels. Their triple invariants differ by approximately a factor of $370$ on the finer sampled grid, primarily due to the Coulomb potential's variation along joint density levels. 

\item[Conclusions] The pair invariant measures variation of one field over the levels of another, while the factorized triple measures the third field's variation along their intersections. Together, these diagnostics reveal complementary aspects of relative field geometry in isolated nuclei, nuclear collisions, and periodic nuclear pasta, without requiring a preferred origin or axis. A resolved nonzero triple excludes a common symmetry axis, but zero does not establish one. Collision refinement tests extend to $110$ fm/$c$, while the periodic-pasta results require further convergence studies.

\end{description}
\end{abstract}

\maketitle

\section{Introduction}

Proton and neutron densities in a mean-field solution need not have the same spatial structure. Similar radii or quadrupole moments do not settle a local question: is one density constant on a surface where the other is fixed? With a third field, such as the Coulomb potential, a second question arises: does it vary along the curves where the two density surfaces intersect?

This work uses differential forms to answer these two questions separately. The squared cross product of two gradients measures variation along a level surface. Its integral gives a weighted measure of their misalignment, and the coarea formula locates that signal by density level. The squared triple product adds the variation of a third field along the intersection curves. We factor its normalized integral into a weighted pair term and a directional term. This separates changes in density misalignment from changes in the third field's contribution.

Cross-gradients already compare structures in joint geophysical inversion~\cite{GallardoMeju2003,GallardoMeju2004,GallardoMeju2011}, multimodal imaging~\cite{HaberModersitzki2006}, and inverse-problem regularization~\cite{Crestel2018}. They also enter baroclinic vorticity generation~\cite{Pedlosky1987} and the Biermann battery~\cite{Biermann1950}. Their zeros are related to the Jacobi set~\cite{EdelsbrunnerHarer2004}; joint level curves and families of these curves are studied in bivariate visualization~\cite{Carr2015}. In nuclear physics, joint contour nets have been used to analyze the topology of proton and neutron level sets in fission~\cite{Duke2012}. Our contribution is a metric comparison for nuclear mean fields: measuring tangential variation, locating it on level surfaces, and separating the two contributions to the triple integral.

Multipole expansions can recover this information when enough components of both fields are retained. Minkowski functionals describe the morphology of individual density surfaces~\cite{Schuetrumpf2013,Schneider2013}. The quantities used here compare fields directly, without choosing an origin, an axis, or a multipole truncation.

Three applications test the argument, followed by a preliminary periodic-pasta example. Magnetized Wigner--Seitz cells show differences that selected low-order moments leave unresolved. Static nuclei locate the density differences on their surfaces and test the triple against a common symmetry axis. A collision shows why the factorization matters: a long-range potential can change the triple while the weighted density-pair term remains nearly constant. The periodic example compares a slab and a connected network at the same density and composition, with spatial convergence still under study. These are tests of relative field geometry; the invariants do not determine neutron-skin thickness or quantify triaxiality.

Section~\ref{sec:geometry} defines the diagnostics and their level-surface interpretation. Section~\ref{sec:numerics} describes their evaluation and numerical checks. Sections~\ref{sec:cells}--\ref{sec:collision} present the three applications, and Sec.~\ref{sec:pasta} gives the preliminary periodic-pasta comparison. Section~\ref{sec:scope} compares the diagnostics with other measures and discusses their limits.

\section{Relative geometry of the fields}
\label{sec:geometry}

\subsection{A weighted measure of pair misalignment}

Let $f_a$ and $f_b$ be scalar fields on a domain with metric $g$ and volume element $dV$. Their differentials are 1-forms, with $|df|^2=g^{ij}\partial_i f\,\partial_j f$. We define
\begin{align}
 I_{ab}&=\int|df_a\wedge df_b|^2\,dV, \label{eq:I}\\
 S_{ab}&=\frac{I_{ab}}{\int|df_a|^2|df_b|^2\,dV}\in[0,1], \label{eq:S}\\
 A_{ab}(\bm r)&=\frac{|df_a\wedge df_b|^2}{|df_a|^2|df_b|^2}
 =\sin^2\angle(\nabla f_a,\nabla f_b). \label{eq:A}
\end{align}
The metric gives $|df_a\wedge df_b|^2=|df_a|^2|df_b|^2-\langle df_a,df_b\rangle^2$. Thus $S_{ab}$ is the mean of $A_{ab}$ weighted by $|df_a|^2|df_b|^2$. It is defined when the integrated denominator is positive; the local quantity $A_{ab}$ requires both gradients to be nonzero. Regions of small gradients contribute little to the integrated measure, even when a local angular map there is dominated by numerical noise.

For the same fields and domain, $I_{ab}$ is twice the cross-gradient penalty of Crestel et al.~\cite{Crestel2018}. Their normalized variant integrates a regularized local angular quantity. Equation~\eqref{eq:S} instead normalizes after integration, retaining the gradient weights. This choice emphasizes the parts of the fields where both vary appreciably.

The quantities are independent of the coordinates used to describe the domain, provided the metric and volume element are transformed with them. A physical deformation of the fields or domain can change them. For continuously differentiable fields, $I_{ab}=0$ precisely when $df_a\wedge df_b$ vanishes throughout the domain. Wherever $\nabla f_a\neq0$, this means that $f_b$ is locally a function of $f_a$; separate regions need not share one global relation. Parallel and antiparallel gradients both give zero.

This null case sets the scope of the diagnostic. Two spherical densities have $S_{ab}=0$ even if their radii or diffuseness differ. More generally, a density that is a function of another gives zero whatever their surface offsets. A nonzero pair signal measures disagreement of local gradient directions, not the size of a neutron skin.

The local angle $A_{ab}$ is unchanged by replacing either field with $g(f)$ when $g'(f)\neq0$. The integrated quantities generally change because their weights change. Logarithmic densities give the tails greater weight than linear densities. On a mesh, the chain rule $\nabla g(f)=g'(f)\nabla f$ preserves the local angular invariance, whereas differentiating the sampled transformed field need not. We use linear densities throughout the main analysis.

\subsection{Where the pair signal lies: level surfaces}
\label{sec:levels}

On a regular level surface $\Sigma_c=\{f_a=c\}$, the unit normal is $\hat{\bm n}_a=\nabla f_a/|\nabla f_a|$. Projecting the other gradient into the tangent plane gives $\nabla_\Sigma f_b=(\mathbf 1-\hat{\bm n}_a\hat{\bm n}_a^T)\nabla f_b$ and
\begin{equation}
 |df_a\wedge df_b|^2=|\nabla f_a|^2|\nabla_\Sigma f_b|^2. \label{eq:tangential}
\end{equation}
The pair signal therefore measures the variation of $f_b$ along a surface on which $f_a$ is fixed. On regular levels it vanishes at critical points of that restriction, giving the regular-level interpretation of the Jacobi set~\cite{EdelsbrunnerHarer2004}.

The coarea formula $\int h|\nabla f_a|\,dV=\int dc\int_{\Sigma_c}h\,dA$~\cite{Federer1969,Howard1994} resolves the integral into these surfaces:
\begin{equation}
 I_{ab}=\int dc\,\mathcal E_{ab}(c),\qquad
 \mathcal E_{ab}(c)=\int_{\Sigma_c}|\nabla f_a||\nabla_\Sigma f_b|^2\,dA.
 \label{eq:coarea}
\end{equation}
Replacing $|\nabla_\Sigma f_b|$ by $|\nabla f_b|$ defines the corresponding denominator profile $\mathcal D_{ab}(c)$. Where it is positive, $s_{ab}(c)=\mathcal E_{ab}(c)/\mathcal D_{ab}(c)$ is the misalignment at level $c$. Its mean weighted by $\mathcal D_{ab}$ is $S_{ab}$. The two profiles answer different questions: $\mathcal E_{ab}$ locates the levels carrying the integrated signal, while $s_{ab}$ compares their local geometry without the total amplitude of that signal.

\subsection{What a third field adds: intersection curves}
\label{sec:3d}

In three dimensions, two regular level surfaces with independent normals intersect along curves $\gamma(c_a,c_b)$. Their tangent direction is
\begin{equation}
 \bm b_{ab}=\nabla f_a\times\nabla f_b, \label{eq:b}
\end{equation}
the vector corresponding to the Hodge dual of $df_a\wedge df_b$. Its norm is the Jacobian of $F=(f_a,f_b):\mathbb R^3\to\mathbb R^2$. The coarea formula for this map~\cite{Federer1969,Howard1994} gives
\begin{equation}
 I_{ab}=\int dc_a\,dc_b\int_{\gamma(c_a,c_b)}|\bm b_{ab}|\,d\ell. \label{eq:curves}
\end{equation}
The integral sums the lengths of the joint level curves, weighted by the gradient magnitudes and the sine of the angle between them. These curves are the fibers of $F$; families of fibers over curves in its value plane form fiber surfaces~\cite{Carr2015}.

A third field $f_c$ can vary along a joint level curve even though $f_a$ and $f_b$ remain constant there. Its derivative along the curve is $\partial_\ell f_c=\nabla f_c\cdot\hat{\bm b}_{ab}$. The squared triple product measures this variation:
\begin{align}
 I_{abc}&=\int|df_a\wedge df_b\wedge df_c|^2\,dV \nonumber\\
 &=\int(\nabla f_c\cdot\bm b_{ab})^2\,dV
 =\int|\bm b_{ab}|^2(\partial_\ell f_c)^2\,dV. \label{eq:triple}
\end{align}
Writing $\partial_\ell f_c=|\nabla f_c|\cos\alpha$, and defining $g=|\nabla f_a|^2|\nabla f_b|^2$, $j=|\bm b_{ab}|^2$, and $w=|\nabla f_c|^2$, yields the exact factorization
\begin{gather}
 S_{abc}=\frac{I_{abc}}{\int gw\,dV}
 =\frac{\int jw\cos^2\alpha\,dV}{\int gw\,dV}
 =S^{(c)}_{ab}T_{ab,c}, \label{eq:factor}\\
 S^{(c)}_{ab}=\frac{\int jw\,dV}{\int gw\,dV},\qquad
 T_{ab,c}=\frac{\int jw\cos^2\alpha\,dV}{\int jw\,dV}. \nonumber
\end{gather}
Both factors lie in $[0,1]$ when their denominators are positive. At points where $j$ or $w$ vanishes, the angular quantity need not be defined; its weighted numerator is the squared scalar product in Eq.~\eqref{eq:triple} and remains well defined.

$S^{(c)}_{ab}$ measures the pair misalignment with the additional weight $|\nabla f_c|^2$. $T_{ab,c}$ measures the alignment of the third gradient with the intersection curves, weighted by $jw$. A small triple can have two different origins: the first two fields may be nearly aligned, or the third may vary little along their intersections. Factoring the triple distinguishes these contributions. The directional factor still samples the regions selected by the pair through $j$, so it is not a diagnostic of $f_c$ alone. A change in $T_{ab,c}$ can reflect changes in angles, weights, or both.

\subsection{Symmetry and the information in multipoles}

If all three fields share an axis, their gradients lie in meridian planes and the tangent $\bm b_{ab}$ is azimuthal. Where the intersection curves are regular, they are circles about that axis, and $T_{ab,c}=S_{abc}=0$. A resolved nonzero triple therefore excludes a common axis, regardless of its orientation. Zero is inconclusive: for any triaxial field $f$, the fields $f$, $f^2$, and $f^3$ have parallel gradients and a vanishing triple. In that example the pair is also zero, so $T_{ab,c}$ is undefined. Thus only a nonzero triple can rule out a common axis; its magnitude does not quantify triaxiality.

For axial fields the pair has a single component. In any orthogonal meridian coordinates $(u,v)$ with scale factors $h_u,h_v$,
\begin{equation}
 |df_a\wedge df_b|=
 \left|\frac{\partial(f_a,f_b)}{\partial(\varrho,z)}\right|
 =\frac{|\partial_u f_a\partial_v f_b-\partial_v f_a\partial_u f_b|}{h_uh_v}.
 \label{eq:jac}
\end{equation}
Here $\varrho$ is the distance from the axis. Over a planar meridian region $D$, the signed integral $\int_Ddf_a\wedge df_b$ is the oriented image area counted with multiplicity in the $(f_a,f_b)$ plane. Opposite orientations can cancel. The squared norm in $I_{ab}$ avoids this cancellation and measures where the two gradients are linearly independent.

When one field is nearly spherical, $f_b\simeq f_b(r)$, the pair norm reduces to $|r^{-1}\partial_\theta f_a\,\partial_r f_b|$. A Legendre component $P_l$ then contributes an angular derivative weight proportional to $l(l+1)$. This explains why fine angular structure can matter despite small low-order moments. A complete expansion of both fields retains the same information; the invariants summarize it without requiring the expansion center.

\section{Evaluating and testing the diagnostics}
\label{sec:numerics}

\subsection{Discrete invariants}

We evaluate the diagnostics from fields sampled at the solver nodes $\bm r_k$, with volume weights $w_k$. From their physical gradients we form
\begin{align}
 g_{ab,k}&=|\nabla f_a|_k^2|\nabla f_b|_k^2, \nonumber\\
 j_{ab,k}&=g_{ab,k}-(\nabla f_a\cdot\nabla f_b)_k^2
 =|df_a\wedge df_b|_k^2, \label{eq:lagrange}
\end{align}
by Lagrange's identity, and
\begin{equation}
 t_{abc,k}=[\nabla f_a\cdot(\nabla f_b\times\nabla f_c)]_k^2
\end{equation}
for the triple. In three dimensions we compute $j_{ab,k}$ from the squared components of $\nabla f_a\times\nabla f_b$, avoiding subtraction of nearly equal terms when gradients align. Quadrature gives $I_{ab}=\sum_k w_kj_{ab,k}$ and $S_{ab}=\sum_k w_kj_{ab,k}/\sum_k w_kg_{ab,k}$, with analogous sums for the triple and its factors. A local map $A_{ab,k}$ is displayed only where $\sqrt{g_{ab,k}}$ exceeds $3\%$ of its maximum. Section~\ref{sec:val} and the grid comparisons test the accuracy of this evaluation.

\subsection{Axial meshes}

The axial analysis uses physical gradient components in the meridian plane and the exact finite-volume cell weights. Second-order differences act on the sampled fields; analytic scale factors convert them into physical components. The half-cell covers $z\geq0$, and reflection symmetry supplies the other half. The coordinate formulas and tests are given in Appendix~\ref{app:axial}.

\subsection{Cartesian meshes}
\label{sec:cartesian}

Three-dimensional Hartree--Fock codes such as Sky3D~\cite{Maruhn2014} use uniform Cartesian boxes with spacings $\Delta x$, $\Delta y$, $\Delta z$ of about $1$ fm and represent derivatives spectrally. We do the same. For a field $f$ on $n_x\times n_y\times n_z$ points,
\begin{equation}
 \partial_xf=\mathcal F^{-1}\bigl[ik_x\,\mathcal F[f]\bigr],\qquad k_x=\frac{2\pi m}{n_x\Delta x},
\end{equation}
with $m=-n_x/2+1,\dots,n_x/2-1$ and the Nyquist component $m=n_x/2$ set to zero, and analogously for $y$ and $z$; $\mathcal F$ is the discrete Fourier transform. The quadrature weight is $\Delta x\Delta y\Delta z$ at every node. This assumes that the field is periodic in the box, which holds to high accuracy for densities that vanish at the boundary but not for the Coulomb potential, which decays as $1/r$. For the potential $\mathrm{erf}(r/a)/r$ of a Gaussian charge ($a=2.5$ fm) in a $30$ fm box with $1$ fm spacing, the relative $L^2$ error of the spectral gradient in the shell $1<r<7$ fm is $\coulFFTc$ when the charge is at the center of the box. When the charge is displaced by $3$ or $6$ fm, the periodic extension becomes discontinuous and the error grows to $\coulFFTthree\%$ or $\coulFFTsix\%$, independently of the spacing. Fourth-order centered differences,
\begin{equation}
 \partial_xf\big|_i\simeq\frac{-f_{i+2}+8f_{i+1}-8f_{i-1}+f_{i-2}}{12\Delta x},
\end{equation}
with second-order differences at the two outermost nodes, give $\coulFDfour\%$ wherever the charge is ($\coulFDfourHalf\%$ with $0.5$ fm), against $\coulFDtwo\%$ for second-order differences. For isolated systems we use spectral derivatives for densities and fourth-order differences for the Coulomb potential. For the periodic pasta of Sec.~\ref{sec:pasta}, the Coulomb potential is periodic and all three fields are differentiated spectrally.

For an anisotropic Gaussian pair at $1$ fm spacing, second-order finite differences give $I_{ab}$ with an error of $\errFD\%$ and spectral derivatives with an error of $\errFFT$. Smooth Gaussians do not calibrate the error for nuclear surfaces, however. For two deformed Fermi distributions with diffuseness $0.50$ and $0.55$ fm sampled on the grid, $S_{pn}=\fermiOne$, $\fermiEight$, $\fermiSix$, and $\fermiHalf$ with spacings of $1$, $0.8$, $0.6$, and $0.5$ fm: at $1$ fm the sampled functions have spectral content beyond the Nyquist frequency, and the resulting error dominates their small misalignment. The resolution error of self-consistent densities must therefore be assessed by recomputing the solutions on finer grids (Sec.~\ref{sec:grid}).

\subsection{Level decomposition and surface maps}
\label{sec:levelnum}

Binning the mesh nodes by $f_a$ gives $\mathcal E_{ab}(c_i)\Delta c=\sum_k w_kj_{ab,k}$ for $f_a(\bm r_k)\in[c_i,c_i+\Delta c)$. This is an exact partition of the discrete quadrature. Its interpretation as a surface integral follows from coarea, which also explains the relation between weighted histograms and isosurface statistics~\cite{Scheidegger2008}.

Narrow bins contain few nodes. We reduce their sampling noise by zero-padding the Fourier coefficients of both densities to a grid spacing of about $0.2$ fm, computing the gradients there, and smoothing $\mathcal E_{ab}$ and $\mathcal D_{ab}$ with a Gaussian of width $\sigma=0.04 n_0Z/A$ in $c$. This interpolation reproduces the spectral representation of the original box; it adds no physical resolution. Section~\ref{sec:grid} tests the original fields with finer self-consistent solutions, and Sec.~\ref{sec:surfacecontrast} tests the profile dependence on smoothing and interpolation spacing.

For the isolated-nucleus surface maps we interpolate fields and gradients by cubic splines onto a level surface located along rays from the center of mass (Sec.~\ref{sec:shape}). Contours of a second field on that surface give its intersection curves.

\subsection{Legendre decomposition}

For the comparison with multipole expansions in a spherical cell (Sec.~\ref{sec:beyond}), a density is expanded at each radial node as $n(r_i,\theta)=\sum_{l\ \mathrm{even}}^{l_{\max}}n_l(r_i)P_l(\cos\theta)$ by a least-squares fit to its values at the $N_\theta$ angular nodes. With $N_\theta=28$ nodes on $0<\theta<\pi/2$ we use $l_{\max}=20$; higher orders, amplified by the factor $l(l+1)$ in the gradients, are dominated by aliasing. The invariant of the truncated series is computed with the same discrete method.

\subsection{Rotations and shape measures}
\label{sec:shape}

A computed density is rotated relative to the grid by evaluating its trigonometric interpolant,
\begin{equation}
 f(\bm r)=\sum_{\bm k}\hat f_{\bm k}\,e^{i\bm k\cdot(\bm r-\bm r_0)},
\end{equation}
at the points $R^{T}(\bm r_k-\bm r_c)+\bm r_c$, where $\hat f_{\bm k}$ are the discrete Fourier coefficients, $R$ is the rotation, and $\bm r_c$ the center of mass. Points whose source lies outside the box are set to zero, where the densities vanish; the series would otherwise return the value of a periodic image. The interpolation is exact at the original nodes.

To interpret the invariants of the nuclei we use two measures that do require an axis. The principal axes are those of the quadrupole tensor of the total density about its center of mass, and the symmetry axis is the one with the largest eigenvalue in magnitude. The axial part $\bar f$ of a field is the average of $M=24$ copies rotated about this axis by $2\pi j/M$, which removes all azimuthal components except multiples of $M$; $\delta_f=\|f-\bar f\|/\|f\|$ measures the departure from axial symmetry. Isodensity surfaces are located along rays from the center of mass by cubic-spline interpolation: $R_q(\theta,\phi)$ is the outermost radius where $n_q=\lambda\,n_0\,Q/A$, with $n_0=0.16\,\mathrm{fm}^{-3}$ and $Q=N$ or $Z$. These reference levels are fixed by the composition of the nucleus, not by the local bulk density of each species. We use $\lambda=0.1$ to $0.8$; at higher levels the reference approaches the interior neutron density of $^{34}$Mg. Each surface is characterized by $R_q$ at the tips ($\theta=0$) and on the equator ($\theta=\pi/2$, averaged over $\phi$).

\subsection{Validation and resolution limits}
\label{sec:val}

Analytic tests check the metric factors, the null cases, orthogonal gradients, and convergence of the pair integral on curvilinear meshes (Appendix~\ref{app:axial}). They separate errors in the evaluation of a known field from errors in the self-consistent solution. On Cartesian grids, three anisotropic Gaussians with mutually misaligned axes give $S_{abc}=\oriTriple$. Evaluating them analytically after rotations by Euler angles $(30^\circ,50^\circ,20^\circ)$ and $(117^\circ,71^\circ,203^\circ)$ changes $I_{ab}$, $S_{ab}$, and $S_{abc}$ by at most $\oriAnaOne$ at $1$ fm and $\oriAnaHalf$ at $0.5$ fm.

For $f_a=e^{-r^2/a^2}$ and $f_b=z e^{-r^2/s^2}$, the level profile is $\mathcal E_{ab}(c)=(16\pi/3)c r_c^3 e^{-2r_c^2/s^2}/a^2$, with $r_c=a\sqrt{\ln(1/c)}$. Fourfold refinement from $1$ and $0.6$ fm grids, followed by the same smoothing of the discrete and analytic profiles, gives errors of $\coareaOne\%$ and $\coareaSix\%$. The triple factorization holds to rounding. Three axial Gaussians with a common axis tilted relative to the grid give $T_{ab,c}=\Tfloor$ at $1$ fm.

These checks establish the implementation for smooth analytic fields. The deformed Fermi profiles above show why nuclear surfaces need an additional test: a small angular signal can be obscured by sampling error. For the nuclear results we therefore compare newly computed solutions at several spacings and box sizes, rather than using interpolated fields as evidence of physical convergence.

\section{Density differences missed by selected moments}
\label{sec:cells}

\subsection{Magnetized cell configurations}

The first test compares fields in \Nsol{} converged relativistic mean-field solutions of a spherical Wigner--Seitz cell. We use NL3~\cite{Lalazissis1997}, radius $16.73$ fm, mean baryon density $0.019\,\mathrm{fm}^{-3}$, and proton fraction $0.3$. Neutrons and electrons are treated in the Thomas--Fermi approximation; Dirac protons couple minimally to $\bm B=B\hat z$ in a confined basis. Appendix~\ref{app:dirac} describes the Dirac solver, its approximations and parameters, and the basis and mesh checks for these solutions. The solutions cover $B=0$ to $10^{18}$ G, quadrupole constraints, and two starting bases. For each field we use the unconstrained solution of lower energy; at $B=0$ the two starting bases converge to distinct oblate and prolate minima.

The neutron density responds to fields sourced by both species and is not spherical. In the unconstrained solutions the $L^2$ norm of its non-monopole part, $\|n-\bar n\|/\|\bar n\|$ with $\bar n(r)$ the angular average, is $\nnAngZero\%$ at $B=0$ and $\nnAngSeven\%$ at $7\times10^{17}$ G, compared with $\npAngZero\%$ and $\npAngSeven\%$ for protons. Both densities therefore contribute to the pair signal.

\subsection{Which fields share a structure?}

\begin{figure}[tbp]
 \includegraphics[width=\columnwidth]{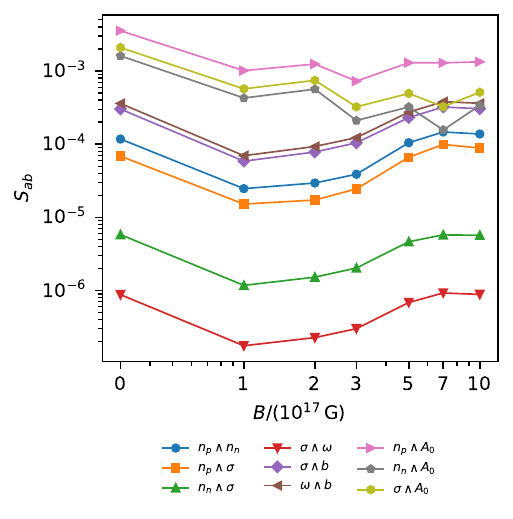}
 \caption{Normalized invariants $S_{ab}$ of nine pairs of fields against the magnetic field, for the unconstrained solution of lower energy at each field. $b$ is the isovector meson field.}
 \label{fig:pairs}
\end{figure}

The nine pairs in Fig.~\ref{fig:pairs} separate according to how closely their gradient directions agree. The scalar and vector-isoscalar fields predominantly follow the baryon distribution and give $S_{\sigma\omega}\simeq\Ssw$. The neutron density is also nearly aligned with the scalar field ($\Snsig$). Pairs involving the proton density or the isovector field $b$, sourced by $n_p-n_n$, are larger by one to two orders of magnitude. Coulomb pairs have the largest values, consistent with a potential whose long range samples the cell as well as the cluster.

Most pairs without the Coulomb potential share a minimum at $1$--$2\times10^{17}$ G and a rise above $3\times10^{17}$ G. Coulomb pairs instead decrease from $B=0$ and then remain roughly constant.

\begin{figure}[tbp]
 \includegraphics[width=\columnwidth]{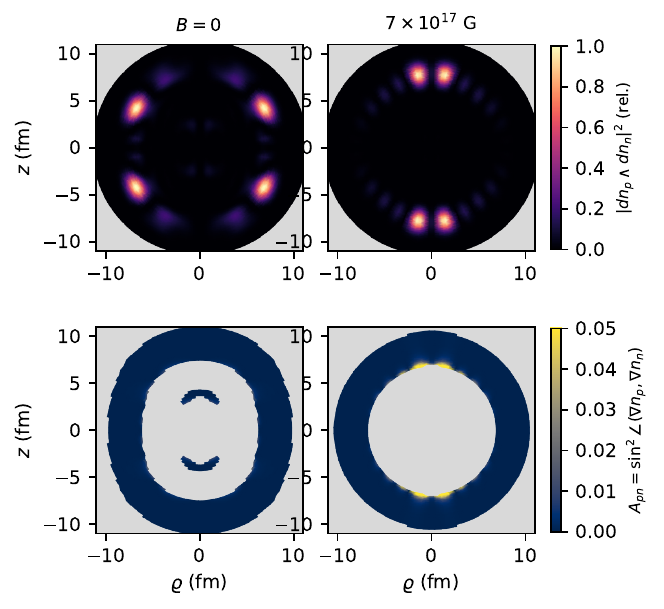}
 \caption{Top: integrand $|dn_p\wedge dn_n|^2$ of $I_{pn}$, normalized to its maximum. Bottom: local misalignment $A_{pn}=\sin^2\angle(\nabla n_p,\nabla n_n)$, shown where $|dn_p||dn_n|$ exceeds $3\%$ of its maximum (gray elsewhere). Unconstrained solutions at $B=0$ and $7\times10^{17}$ G.}
 \label{fig:maps}
\end{figure}

The local signal identifies where the densities disagree (Fig.~\ref{fig:maps}). At $B=0$, $|dn_p\wedge dn_n|^2$ forms four surface lobes at intermediate angles, consistent with a quadrupole difference. At $7\times10^{17}$ G the signal shifts toward the poles. There the local map also shows the largest angles between the gradients.

\subsection{Comparison with multipole moments}
\label{sec:beyond}

\begin{figure*}[tp]
 \includegraphics[width=0.88\textwidth]{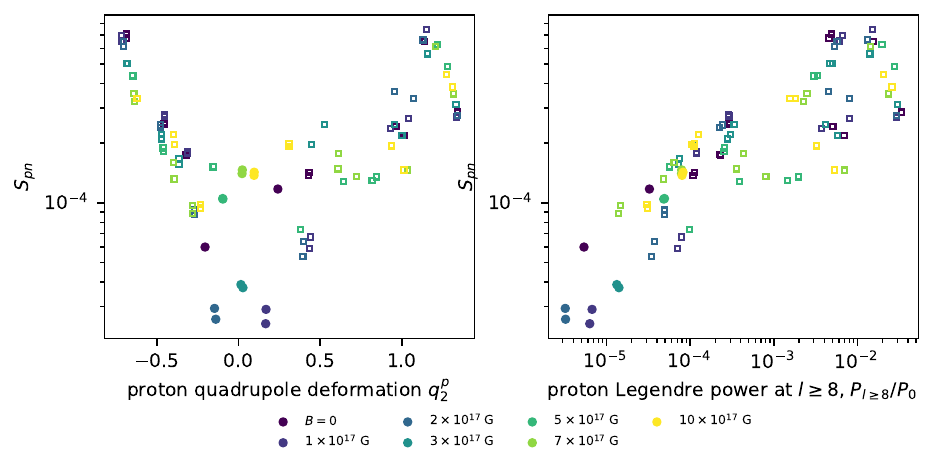}
 \caption{$S_{pn}$ against the proton quadrupole deformation (left) and against the ratio of the proton Legendre power at $l\geq8$ to the monopole power (right). Filled circles: unconstrained solutions; open squares: solutions with a quadrupole constraint. Colors give the field.}
 \label{fig:beyond}
\end{figure*}

Comparable low-order moments do not imply comparable pair geometry. Among solutions at different fields whose proton quadrupole deformations differ by at most $\qTol$ (Fig.~\ref{fig:beyond}, left), $S_{pn}$ differs by up to a factor of $\qContrast$. Here $q_2=Q_{20}/(N\langle r^2\rangle)$ and $q_4=Q_{40}/(N\langle r^2\rangle^2)$, with $Q_{20}=\int n\,(2z^2-\varrho^2)\,dV$ and $Q_{40}=\int n\,r^4P_4(\cos\theta)\,dV$, evaluated after subtracting the mean density of the outermost mesh layer; $q_2$ is not the $\beta_2$ convention. The nearly spherical selected solutions at $3\times10^{17}$ G and $7\times10^{17}$ G both have small quadrupole deformations ($q_2^p=\qthree$ and $\qseven$), but their hexadecapole deformations differ in sign ($q_4^p=\qfourthree$ and $\qfourseven$), and $S_{pn}=\Spnthree$ and $\Spnseven$.

This difference is compatible with a multipole description. Reconstructing both densities through $l=20$ reproduces $I_{pn}$ with mean ratio $\invSeries$, standard deviation $\invSeriesSd$, and largest deviation $\invSeriesMax$. For unconstrained solutions, $\log S_{pn}$ and the logarithm of the proton power ratio $P_{l\geq8}/P_0$ have correlation coefficient \corrSP{} (Fig.~\ref{fig:beyond}, right). This association is specific to the sampled solutions: the invariant depends on both fields and their radial gradient weights.

A lower-order truncation cannot assign the integral to separate multipoles, since products of the reconstructed fields generate cross terms and cancellations. The pair diagnostic therefore supplies a compact comparison when selected moments are insufficient, while a sufficiently complete expansion of both fields recovers it.

Recomputing the unconstrained solution at $7\times10^{17}$ G with $N_F=24$ oscillator shells instead of $\solNF$ changes $S_{pn}$ by $\chkBasisSeven\%$, and on a $120\times40$ instead of $\solMesh$ mesh by $\chkMeshSeven\%$; at $10^{18}$ G the larger basis changes it by $\chkBasisTen\%$ for the solution from the prolate starting basis, which is not the selected one at this field (Appendix~\ref{app:dirac}). These changes are smaller than the contrasts above.

\section{Density surfaces and a test of common axial symmetry}
\label{sec:nuclei}

\subsection{Static configurations}

We next compare static configurations of $^{20}$Ne, $^{24}$Mg, $^{34}$Mg, $^{76}$Ge, and $^{128}$Xe. They were computed with Sky3D v1.2~\cite{Maruhn2014,Sky3D2024}, SV-bas~\cite{Kluepfel2009}, no pairing, and cubic boxes of $26$--$36$ fm at $1$ fm spacing. The recorded code commit is \texttt{be42efc7} (the full identifier is in the data manifest). Sky3D imposes no spatial symmetry. The three light configurations are prolate with their axis along $x$; $^{76}$Ge and $^{128}$Xe retain triaxial shapes after initialization from triaxial oscillator states.

We did not compare alternative initializations. The calculation without pairing tests the diagnostics on these configurations and does not predict the physical neutron skin. Table~\ref{tab:nuclei} gives the density moments, invariants, and the independent axial-departure measure $\delta_n$.

\begin{table*}[tp]
 \caption{Static Skyrme Hartree--Fock configurations (SV-bas, no pairing, $1$ fm spacing). Quadrupole deformations of protons and neutrons, triaxiality angle of the neutron density, difference of the rms radii, invariants of the proton and neutron densities and of the proton density and the Coulomb potential $W$, triple invariant of the two densities and $W$, its factor $T_W=T_{pn,W}$ of Eq.~\eqref{eq:factor}, and the relative non-axial part $\delta_n$ of the neutron density (Sec.~\ref{sec:shape}).}
 \label{tab:nuclei}
 \begin{ruledtabular}
 \begin{tabular}{lrrrrrrcccccc}
 nucleus & $Z$ & $N$ & $\beta_2^p$ & $\beta_2^n$ & $\gamma_n$ ($^\circ$) & $r_n-r_p$ (fm) & $S_{pn}$ & $S_{pW}$ & $S_{npW}$ & $T_W$ & $\delta_n$\\
 $^{20}$Ne & 10 & 10 & 0.538 & 0.522 & 0.0 & -0.031 & $1.5\times10^{-5}$ & $2.5\times10^{-2}$ & $2.1\times10^{-9}$ & $1.6\times10^{-4}$ & $2.3\times10^{-3}$ \\
$^{24}$Mg & 12 & 12 & 0.526 & 0.514 & 0.0 & -0.032 & $1.8\times10^{-5}$ & $2.4\times10^{-2}$ & $1.8\times10^{-9}$ & $1.1\times10^{-4}$ & $2.7\times10^{-3}$ \\
$^{34}$Mg & 12 & 22 & 0.347 & 0.325 & 0.0 & +0.392 & $4.5\times10^{-3}$ & $1.4\times10^{-2}$ & $1.8\times10^{-7}$ & $4.0\times10^{-5}$ & $1.1\times10^{-3}$ \\
\colrule
$^{76}$Ge & 32 & 44 & 0.196 & 0.179 & 8.7 & +0.136 & $8.3\times10^{-4}$ & $5.4\times10^{-3}$ & $7.6\times10^{-7}$ & $9.4\times10^{-4}$ & $2.4\times10^{-2}$ \\
$^{128}$Xe & 54 & 74 & 0.134 & 0.111 & 10.3 & +0.109 & $3.4\times10^{-4}$ & $5.1\times10^{-3}$ & $8.8\times10^{-7}$ & $2.6\times10^{-3}$ & $2.1\times10^{-2}$ \\

 \end{tabular}
 \end{ruledtabular}
\end{table*}

\begin{figure*}[tp]
 \includegraphics[width=\textwidth]{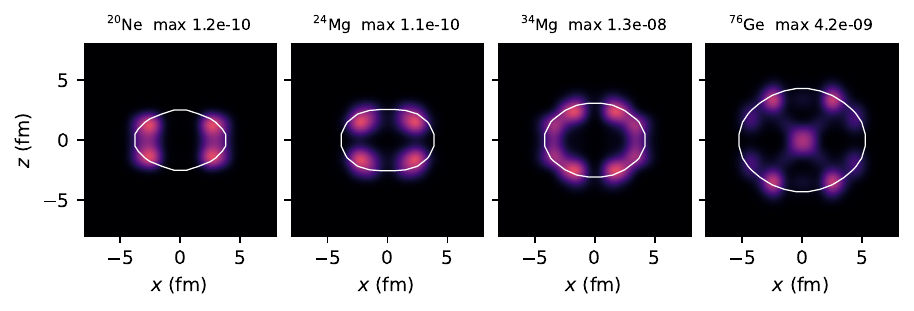}
 \caption{Integrand $|dn_p\wedge dn_n|^2$ (fm$^{-16}$) in the plane through the center of mass that contains the symmetry axis ($x$) of $^{20}$Ne, $^{24}$Mg, and $^{34}$Mg, and the longest principal axis of $^{76}$Ge. The densities are translated by a Fourier phase so that the plane passes exactly through the center of mass. Each panel has its own color scale, with its maximum indicated. White curve: total density $0.08\,\mathrm{fm}^{-3}$.}
 \label{fig:nuclei}
\end{figure*}

\subsection{A contrast distributed over the surface}
\label{sec:surfacecontrast}

The $N=Z$ configurations have nearly aligned proton and neutron gradients, with $S_{pn}\simeq1.5$--$1.8\times10^{-5}$. In $^{34}$Mg, $S_{pn}=\SmgThreeFour$, a factor of \SratioMg{} above $^{24}$Mg despite similar proton and neutron quadrupole deformations within $^{34}$Mg ($0.347$ and $0.325$). Figure~\ref{fig:nuclei} locates the difference: the integrand covers the $^{34}$Mg surface, whereas in the $N=Z$ nuclei it is concentrated near the tips.

\begin{figure}[tbp]
 \includegraphics[width=\columnwidth]{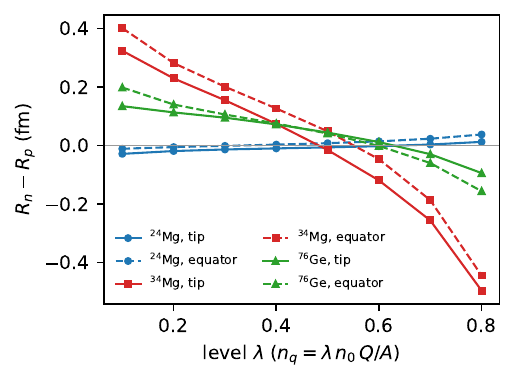}
 \caption{Difference between the neutron and proton surface radii at the reference levels $n_q=\lambda n_0Q/A$, at the tips (solid) and on the equator (dashed), for $^{24}$Mg, $^{34}$Mg, and $^{76}$Ge ($1$ fm spacing).}
 \label{fig:surfaces}
\end{figure}

The density surfaces give a direct description of this contrast (Fig.~\ref{fig:surfaces}). In the $N=Z$ nuclei, radii at matched reference levels differ by at most $\drNZmax$ fm, and the thickness $R_q(0.3)-R_q(0.7)$, averaged between tip and equator, is $\tNZmin$--$\tNZmax$ fm for both species. In $^{34}$Mg it is $\tnMg$ fm for neutrons and $\tpMg$ fm for protons. Accordingly, $R_n-R_p$ changes from $\drThreeMg$ fm at $\lambda=0.3$ to $\drSevenMg$ fm at $\lambda=0.7$, even though the rms-radius difference is $\skinMg$ fm.

Surface displacement and diffuseness are distinct contributions to a neutron skin; valence orbitals can affect them differently~\cite{Warda2014}. The surfaces measured here show that the $^{34}$Mg difference is dominated by the greater neutron diffuseness. Its offset also depends on direction: tip and equatorial offsets differ by $\anMgMin$--$\anMgMax$ fm at $\lambda=0.3$--$0.7$, compared with at most $\anNZmax$ fm in the $N=Z$ nuclei. Direction-dependent skins have also been studied through heavy-ion observables~\cite{Liu2023}.

The pair signal measures the angular disagreement associated with these surfaces; the radii and thicknesses require the separate surface measurements above. In $^{76}$Ge and $^{128}$Xe, $S_{pn}=\SpnGe$ and $\SpnXe$, respectively, and neutron surfaces are thicker by $\dtTriMin$ fm.

Changing the field weights changes the question being asked. With logarithmic densities evaluated by the chain rule and a cut on both densities, $S_{pn}$ varies by up to a factor of $\logVar$ as the cut moves from $10^{-4}$ to $10^{-2}\,\mathrm{fm}^{-3}$. The tails then dominate. Only the contrast between $^{34}$Mg and the $N=Z$ configurations persists across these cuts, with ratios $\logRatioMin$--$\logRatioMax$. We therefore base the level analysis on linear densities.

\begin{figure}[tbp]
 \includegraphics[width=\columnwidth]{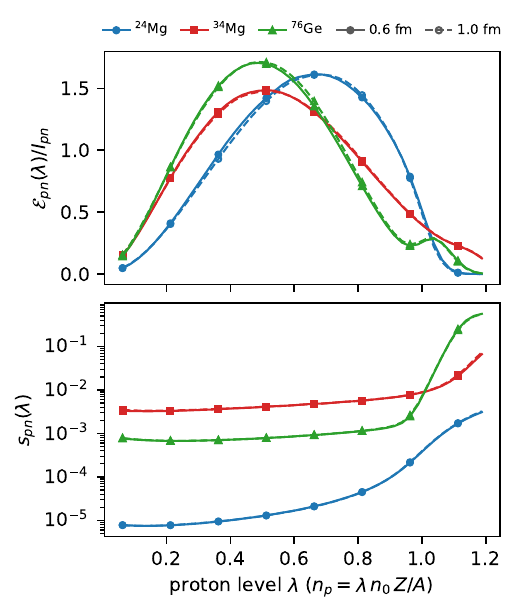}
 \caption{Decomposition of $I_{pn}$ over the proton level surfaces $n_p=\lambda n_0Z/A$ [Eq.~\eqref{eq:coarea}] for $^{24}$Mg, $^{34}$Mg, and $^{76}$Ge, from the solutions with $0.6$ fm (solid) and $1$ fm (dashed) spacing. Top: $\mathcal E_{pn}$ per unit $\lambda$, normalized to $I_{pn}$. Bottom: level misalignment $s_{pn}=\mathcal E_{pn}/\mathcal D_{pn}$.}
 \label{fig:levels}
\end{figure}

Resolving the pair integral by proton level clarifies whether one layer accounts for the contrast. Figure~\ref{fig:levels} uses $c=\lambda n_0Z/A$. Most of $I_{pn}$ lies between $\lambda\approx0.2$ and $0.9$, with profile maxima at $\lvPeakMgTwo$ in $^{24}$Mg, $\lvPeakMgThree$ in $^{34}$Mg, and $\lvPeakGe$ in $^{76}$Ge. The main surface carries the signal. Interior levels above $\lambda\approx1$ can have large $s_{pn}$, but their weak gradients give little integrated weight.

The level misalignment $s_{pn}$ of $^{34}$Mg exceeds that of $^{24}$Mg at every proton level from $\lambda=0.1$ to $0.9$, by factors of $\lvRatioMin$--$\lvRatioMax$. The contrast is distributed over the surface rather than confined to a selected layer. Profiles from the $1$ and $0.6$ fm solutions agree to $\lvDev\%$ of their maximum.

This interpretation survives changes in the profile estimator. Halving or doubling $\sigma$ shifts maxima by at most $\lvSigPeak$ in $\lambda$ and changes the integral by at most $\lvSigInt\%$. It changes $s_{pn}$ by at most $\lvSigSeight\%$ for $0.1\leq\lambda\leq0.8$, but up to $\lvSigS\%$ near $0.9$, where the profile rises steeply. Refinement to $0.15$ instead of $0.2$ fm changes $s_{pn}$ by less than $\lvUpS\%$. Across these variants, the $^{34}$Mg/$^{24}$Mg contrast remains $\lvRatioVarMin$--$\lvRatioVarMax$.

\begin{figure}[tbp]
 \includegraphics[width=\columnwidth]{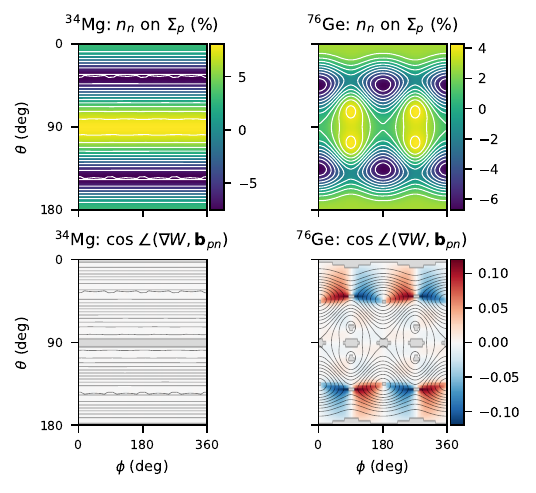}
 \caption{Intersection curves on the proton level surface $\Sigma_p$ ($n_p=0.5\,n_0Z/A$) of $^{34}$Mg and $^{76}$Ge ($0.6$ fm spacing), parametrized by the polar angle $\theta$ from the symmetry (longest) axis and the azimuth $\phi$. Top: $n_n$ on $\Sigma_p$, relative to its mean; the contours are the intersections with neutron level surfaces. Bottom: cosine of the angle between $\nabla W$ and the curve tangent $\bm b_{pn}$, shown where $|\bm b_{pn}|$ exceeds $10\%$ of its maximum on the surface, with the same contours.}
 \label{fig:intersections}
\end{figure}

\subsection{Separating pair geometry from the symmetry test}
\label{sec:triple}

The triple asks a different question from the pair: does the Coulomb potential vary along the joint density levels? Its raw value includes the amplitude of the pair misalignment. Among the prolate configurations $S_{npW}$ spans two orders of magnitude at $1$ fm (Table~\ref{tab:nuclei}), following $S^{(W)}_{pn}\simeq S_{pn}$; for $^{34}$Mg, $S^{(W)}_{pn}=\SpwMgThree$. This variation cannot be read as an ordering by departure from axial symmetry.

The factorization isolates the directional factor $T_W$. At $1$ fm it lies between $\TWaxMin$ and $\TWaxMax$ in the prolate configurations, compared with $\TWtriMin$--$\TWtriMax$ in $^{76}$Ge and $^{128}$Xe: a separation of at least $\TWsepOne$. The independent axial-departure measure gives the same grouping, $\delta_n=\naxAxMin$--$\naxAxMax$ versus $\naxTriMin$--$\naxTriMax$. This grouping agrees with the common-axis test derived in Sec.~\ref{sec:geometry}.

The intersection curves make the distinction visible (Fig.~\ref{fig:intersections}). At $0.6$ fm, contours of $n_n$ on the proton surface $\lambda=0.5$ are circles of constant $\theta$ in $^{34}$Mg. The Coulomb gradient is perpendicular to them within an rms cosine $\cosMg$, the root mean square over the $(\theta,\phi)$ grid points of the surface where $|\bm b_{pn}|$ exceeds $10\%$ of its maximum; it is neither an area average nor $\sqrt{T_W}$. In $^{76}$Ge, the loops vary with $\phi$ and $\nabla W$ has a component along them: the rms cosine is $\cosGe$, with maxima $\cosGeMax$ in four lobes. These maps show the geometry sampled by $T_W$, while its integrated weights remain those of Eq.~\eqref{eq:factor}.

The separation grows under refinement (Table~\ref{tab:conv}). In the prolate configurations, $S_{npW}$ falls by two to three orders of magnitude between $1$ and $0.6$ fm, $T_W$ reaches $\TWaxSixMin$--$\TWaxSixMax$, and $\delta_n$ falls from about $10^{-3}$ to $\cvNaxAxMin$--$\cvNaxAxMax$. The behavior identifies their coarse-grid departure from axial symmetry as a grid effect. In $^{76}$Ge, $S_{npW}$ changes by $\cvWGeVar\%$, $T_W$ changes from $\TWGeOne$ to $\TWGeSix$, and $\delta_n$ remains near $\cvNaxGe$. At $0.6$ fm, its $T_W$ is about $\TWsepSix$ times the largest prolate value, and its raw triple about $\cvWSep$ times.

\subsection{Grid and box dependence}
\label{sec:grid}

The density-pair contrast is stable in the cases recomputed. We solved $^{20}$Ne, $^{24}$Mg, $^{34}$Mg, and $^{76}$Ge at $0.8$ and $0.6$ fm in boxes of approximately the same size, and the three light nuclei at $1$ fm in boxes enlarged by $8$ fm (Table~\ref{tab:conv}). $^{128}$Xe was not recomputed. Between $1$ and $0.6$ fm, $S_{pn}$ changes by at most $\cvSpn\%$; enlarging the boxes changes it by less than $\cvSpnBox\%$. The $^{34}$Mg/$^{24}$Mg ratio is $\cvRatio$ at $0.6$ fm. Energies change by at most $\cvE$ MeV and quadrupole deformations by at most $\cvBeta$; the $^{24}$Mg and $^{34}$Mg surface thicknesses change by at most $\thickFine$ fm. The Coulomb pair $S_{pW}$ is more sensitive, changing by up to $\cvSpW\%$.

A separate rotation test evaluates the existing representation rather than its physical resolution. Rotating the $^{34}$Mg densities by the Euler angles of Sec.~\ref{sec:val} changes $S_{pn}$ by at most $\oriMgS\%$ and the quadrupole deformations by less than $\oriMgB\%$.

The reported densities and energies refer to the same stored iteration. The light nuclei at $0.6$ fm reached the solver convergence criterion before the next density output. We use the last stored density, up to $\tddGap$ iterations earlier, and the energy from that same iteration. The $^{76}$Ge calculations were continued to $\geLongIter$ iterations; at $0.6$ fm the solver convergence criterion is met after $\geConvIter$. The triaxiality angle settles at $\gamma_n=\geGammaMin$--$\geGammaMax^\circ$ at the three spacings, and between the last two stored densities $S_{pn}$ and $S_{npW}$ change by less than $\geLongDrift$. These are stationary solutions from a single triaxial initialization; other initializations were not compared, pairing is omitted, and $^{128}$Xe was not continued or recomputed.

\begin{table*}[tp]
 \caption{Dependence of the static configurations on the grid spacing $\Delta x$ and the box size $L$: energy, neutron quadrupole deformation, invariants of the proton and neutron densities and of the proton density and the Coulomb potential, triple invariant with the Coulomb potential and its factor $T_W$, and relative non-axial part of the neutron density.}
 \label{tab:conv}
 \begin{ruledtabular}
 \begin{tabular}{lrrrrccccc}
 nucleus & $\Delta x$ (fm) & $L$ (fm) & $E$ (MeV) & $\beta_2^n$ & $S_{pn}$ & $S_{pW}$ & $S_{npW}$ & $T_W$ & $\delta_n$\\
 $^{20}$Ne & 1.0 & 26.0 & -145.59 & 0.522 & $1.457\times10^{-5}$ & $2.5\times10^{-2}$ & $2.1\times10^{-9}$ & $1.6\times10^{-4}$ & $2.3\times10^{-3}$ \\
 & 1.0 & 34.0 & -145.59 & 0.522 & $1.456\times10^{-5}$ & $2.5\times10^{-2}$ & $2.1\times10^{-9}$ & $1.6\times10^{-4}$ & $2.3\times10^{-3}$ \\
 & 0.8 & 25.6 & -145.57 & 0.524 & $1.461\times10^{-5}$ & $2.4\times10^{-2}$ & $1.7\times10^{-10}$ & $1.3\times10^{-5}$ & $5.8\times10^{-4}$ \\
 & 0.6 & 26.4 & -145.58 & 0.524 & $1.460\times10^{-5}$ & $2.4\times10^{-2}$ & $2.1\times10^{-12}$ & $1.6\times10^{-7}$ & $4.6\times10^{-5}$ \\
\colrule
$^{24}$Mg & 1.0 & 28.0 & -183.83 & 0.514 & $1.765\times10^{-5}$ & $2.4\times10^{-2}$ & $1.8\times10^{-9}$ & $1.1\times10^{-4}$ & $2.7\times10^{-3}$ \\
 & 1.0 & 36.0 & -183.83 & 0.514 & $1.765\times10^{-5}$ & $2.4\times10^{-2}$ & $1.9\times10^{-9}$ & $1.1\times10^{-4}$ & $2.7\times10^{-3}$ \\
 & 0.8 & 28.8 & -183.79 & 0.516 & $1.747\times10^{-5}$ & $2.3\times10^{-2}$ & $2.6\times10^{-10}$ & $1.6\times10^{-5}$ & $6.9\times10^{-4}$ \\
 & 0.6 & 27.6 & -183.80 & 0.517 & $1.743\times10^{-5}$ & $2.3\times10^{-2}$ & $6.1\times10^{-12}$ & $3.7\times10^{-7}$ & $5.6\times10^{-5}$ \\
\colrule
$^{34}$Mg & 1.0 & 30.0 & -245.80 & 0.325 & $4.474\times10^{-3}$ & $1.4\times10^{-2}$ & $1.8\times10^{-7}$ & $4.0\times10^{-5}$ & $1.1\times10^{-3}$ \\
 & 1.0 & 38.0 & -245.80 & 0.325 & $4.473\times10^{-3}$ & $1.4\times10^{-2}$ & $1.8\times10^{-7}$ & $4.1\times10^{-5}$ & $1.1\times10^{-3}$ \\
 & 0.8 & 30.4 & -245.80 & 0.324 & $4.411\times10^{-3}$ & $1.4\times10^{-2}$ & $2.0\times10^{-8}$ & $4.5\times10^{-6}$ & $2.5\times10^{-4}$ \\
 & 0.6 & 30.0 & -245.80 & 0.324 & $4.417\times10^{-3}$ & $1.4\times10^{-2}$ & $5.4\times10^{-10}$ & $1.2\times10^{-7}$ & $1.8\times10^{-5}$ \\
\colrule
$^{76}$Ge & 1.0 & 32.0 & -650.18 & 0.179 & $8.322\times10^{-4}$ & $5.4\times10^{-3}$ & $7.6\times10^{-7}$ & $9.4\times10^{-4}$ & $2.4\times10^{-2}$ \\
 & 0.8 & 32.0 & -650.21 & 0.179 & $8.350\times10^{-4}$ & $5.3\times10^{-3}$ & $7.2\times10^{-7}$ & $8.9\times10^{-4}$ & $2.4\times10^{-2}$ \\
 & 0.6 & 32.4 & -650.21 & 0.179 & $8.349\times10^{-4}$ & $5.2\times10^{-3}$ & $7.2\times10^{-7}$ & $8.9\times10^{-4}$ & $2.4\times10^{-2}$ \\

 \end{tabular}
 \end{ruledtabular}
\end{table*}

\section{Separating density and potential contributions before contact}
\label{sec:collision}

\begin{figure}[tbp]
 \includegraphics[width=\columnwidth]{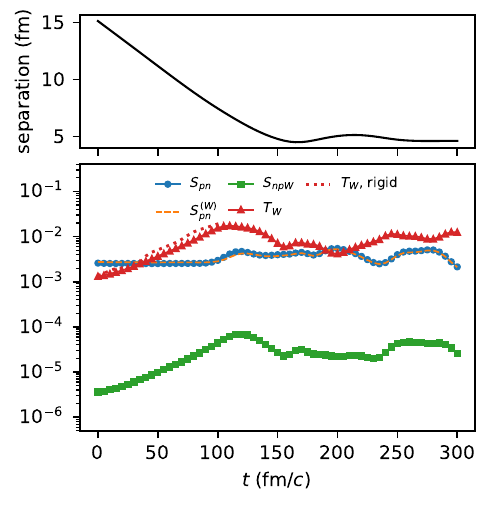}
 \caption{Time-dependent Hartree--Fock collision $^{16}$O\,+\,$^{34}$Mg ($E_{\mathrm{c.m.}}=\collEcm$ MeV, $b=\collB$ fm). Top: distance between the centers of mass of the two halves of the box divided by the plane $z=z_{\mathrm{cm}}$. Bottom: invariant of the proton and neutron densities, triple invariant of the two densities with the Coulomb potential, and its factors $S^{(W)}_{pn}$ and $T_W$ [Eq.~\eqref{eq:factor}], over the whole box. Dotted: $T_W$ for the initial fragments translated rigidly along the trajectory, with the Coulomb potential from Poisson's equation, up to a separation of $7$ fm.}
 \label{fig:collision}
\end{figure}

A collision provides a configuration with two centers and a long-range field that reaches between them. We followed $^{16}$O\,+\,$^{34}$Mg with Sky3D, starting from static solutions, at $E_{\mathrm{c.m.}}=\collEcm$ MeV and impact parameter $\collB$ fm. The grid has $\collGrid$ points at $1$ fm spacing; the time step is $\collDt$ fm/$c$ and the run extends to $\collT$ fm/$c$. All invariants are integrated over the whole box (Fig.~\ref{fig:collision}).

For reference, the plotted separation is the distance between the centers of mass of the two halves divided by $z=z_{\mathrm{cm}}$, with $z$ along the beam. After contact these halves are not physical fragments. The region above total density $0.02\,\mathrm{fm}^{-3}$ becomes connected at $\collConnLow$ fm/$c$, and that above $0.08\,\mathrm{fm}^{-3}$ at $\collConnHigh$ fm/$c$; both remain connected to $\collT$ fm/$c$.

Over $0\leq t\leq\collTpre$ fm/$c$, while the plotted separation remains above $\collSeppre$ fm, $S_{pn}$ stays near $\collSpnPre$ within $\collSpnPreVar\%$. The low-density bridge is first recorded at the end of this interval. For nonoverlapping density gradients the whole-box pair is the mean of the fragment values weighted by $\int|\nabla n_p|^2|\nabla n_n|^2dV$. Here $^{34}$Mg dominates that mean.

During the same interval, $S_{npW}$ rises from $\collWstart$ to $\collWpre$, compared with $\SnpWMg$ for static $^{34}$Mg at the same spacing. Equation~\eqref{eq:factor} assigns this $\collWfacExact$-fold rise to $T_W$, which grows from $\collTWstart$ to $\collTWpre$ (by $\collTWfacExact$); up to $t=\collTdisc$ fm/$c$, the last record with disconnected densities, the two factors are $\collWfacDisc$ and $\collTWfacDisc$. The weighted pair factor $S^{(W)}_{pn}$ changes by at most $\collSwVar\%$. Thus the increase is in the Coulomb-gradient contribution along the density-intersection curves, averaged with weight $jw$, while the weighted pair misalignment remains nearly constant.

The potential of either fragment extends onto the surface of the other, allowing its gradient to have a component along those curves before substantial density overlap. A control separates this geometric effect from dynamical changes of the fields. We split the initial densities at $z=z_{\mathrm{cm}}$, translate each fragment rigidly to the center of the corresponding half of the trajectory at each recorded time, and recompute the direct Coulomb potential by solving Poisson's equation with isolated boundary conditions; at $t=0$ it reproduces the Sky3D potential within $\poissonCheck\%$ where the density exceeds $0.01\,\mathrm{fm}^{-3}$. With this potential in both cases, $T_W$ grows by a factor of $\rigTWfac$ for the rigid fragments and $\dynTWfacP$ along the trajectory up to a separation of $\rigSep$ fm (Fig.~\ref{fig:collision}). The approach of unchanged fragments thus accounts for the rise; the dynamical response of the fields reduces it somewhat. The rigid weighted pair factor varies by up to $\rigSwVar\%$, more than along the trajectory, because the translated tails overlap without self-consistent rearrangement. This interpretation concerns the integrated factors: $T_W$ can change through angles, weights, or both.

After contact $S_{npW}$ reaches $\collWmax$ and then lies between $\collWpostMin$ and $\collWpostMax$. $T_W$ lies between $\collTWpostMin$ and $\collTWpostMax$, while $S_{pn}$ oscillates between $\collSpnPostMin$ and $\collSpnPostMax$ with the connected system's shape vibrations.

Up to $110$ fm/$c$, which covers the approach and the first contact, the calculation is stable under refinement. With a time step of $0.1$ fm/$c$ the invariants change by at most $\convDtMax\%$ at the recorded times. With $0.8$ fm spacing (a $\convGridFine$ grid, starting from static fragments recomputed at that spacing) they change by at most $\convDxMax\%$; the connection times and the last disconnected record are unchanged, and the plotted separation differs by up to $\convSepMax$ fm. Up to that record the triple then rises $\convWfacDiscFine$-fold and $T_W$ $\convTWfacDiscFine$-fold, against $\collWfacDisc$ and $\collTWfacDisc$ at $1$ fm, while $S^{(W)}_{pn}$ varies by at most $\convSwVarFine\%$. The values after contact quoted above, from $150$ fm/$c$ on, rest on the $1$ fm calculation alone.

\section{Relative field geometry in periodic nuclear pasta}
\label{sec:pasta}

Periodic pasta tests the diagnostics in a domain where density surfaces cross the cell boundaries and need not surround a single center. We compare a slab and a connected network at the same mean baryon density and proton fraction. The comparison asks whether similar integrated density misalignment can coexist with different variation of the Coulomb potential along the joint density levels.

\subsection{Periodic configurations and numerical status}

The configurations were calculated with Sky3D~\cite{Maruhn2014,Sky3D2024} using SV-bas, without pairing, at zero temperature and magnetic field. Initial orbitals were obtained from an auxiliary periodic Hamiltonian and filled in spin pairs; the subsequent HF relaxation uses the physical functional without an external guide. No center-of-mass correction is applied to these periodic calculations. Each cubic cell contains $A=648$ nucleons, with $Z=216$, $N=432$, and $\bar n_B=0.04\,\mathrm{fm}^{-3}$. Its side is $L=(A/\bar n_B)^{1/3}=25.303$ fm. Wave functions and the Coulomb solution are periodic in all three directions; a uniform negative background neutralizes the proton charge. Densities and the periodic Coulomb potential $W$ are differentiated spectrally. In particular, the finite differences used for the isolated Coulomb potential in the preceding applications are not used here.

Figures~\ref{fig:pasta_fields} and \ref{fig:pasta_levels} use the $30^3$ fields at iteration 2006 after transfer from the $24^3$ orbitals and renewed HF relaxation, with spacing $L/30=0.8434$ fm. Both configurations meet the operational stability criteria: over the final approximately 200 iterations, the invariants change by at most $0.02\%$, the occupation gaps are positive, and no iteration cycle is detected. Particle numbers are conserved to numerical precision. These checks establish stability of the reported observables, rather than satisfaction of the solver stopping criterion $\mathrm{serr}=10^{-7}$.

The mesh comparison remains preliminary. Between $24^3$ and $30^3$, $S_{pn}$ decreases by $3.3\%$ in the slab and $4.6\%$ in the network; $S_{npW}$ decreases by $26\%$ and $15\%$, respectively. The network comparison uses the continued $24^3$ state at iteration 2512. The triple is therefore not yet mesh-converged. Calculations at $40^3$, independent initializations, and cell-size checks are needed before assigning grid-independent values or physical phase ordering.

\subsection{Similar pair signals, different contributions to the triple}

The slab forms a continuous layer connected through the $x$ and $y$ boundaries. The dense and dilute regions of the network both connect through all three directions. We describe the latter by its observed connectivity rather than identify a specific minimal-surface phase from its initialization.

To locate the integrated signals, define the dimensionless local densities
\begin{align}
 \sigma_{pn}(\bm r)&=\frac{|\nabla n_p\times\nabla n_n|^2}
 {\langle|\nabla n_p|^2|\nabla n_n|^2\rangle_V},\label{eq:pasta_sigma}\\
 \tau_{npW}(\bm r)&=\frac{[(\nabla n_p\times\nabla n_n)\cdot\nabla W]^2}
 {\langle|\nabla n_p|^2|\nabla n_n|^2|\nabla W|^2\rangle_V},\label{eq:pasta_tau}
\end{align}
where $\langle\cdot\rangle_V$ is the cell average. Their cell averages are $S_{pn}$ and $S_{npW}$. Unlike a pointwise angular normalization, these maps retain the gradient weights of the integrated invariants.

\begin{figure*}[t]
 \includegraphics[width=\textwidth]{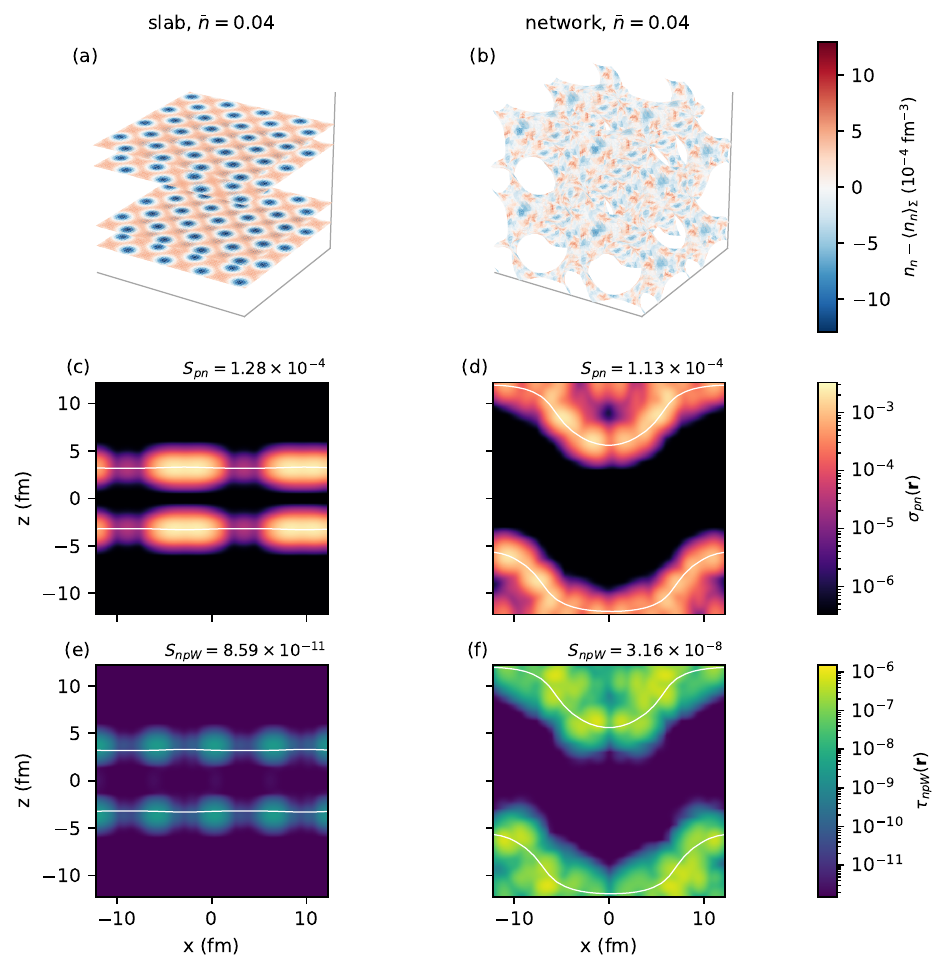}
 \caption{Preliminary periodic slab (left) and network (right) at $\bar n_B=0.04\,\mathrm{fm}^{-3}$, $Y_p=1/3$, on a $30^3$ mesh at iteration 2006 after mesh transfer and HF relaxation. (a, b) Proton level surface $n_p=0.02635\,\mathrm{fm}^{-3}$ over $2\times2\times2$ cells, colored by the deviation of $n_n$ from its area-weighted surface average in one cell. Neutron densities are interpolated to the triangulated surface; colors are unshaded. (c, d) $\sigma_{pn}$ and (e, f) $\tau_{npW}$, defined in Eqs.~\eqref{eq:pasta_sigma} and \eqref{eq:pasta_tau}, on the $xz$ slice with the largest integrated pair contribution for each configuration ($y=4.6$ fm for the slab and $-6.3$ fm for the network); the cell averages $S_{pn}$ and $S_{npW}$ are given above each map. Each row uses a common scale. White contours mark the proton level shown above. Bicubic interpolation is used only for display of the $30^3$ nodal maps; it adds no physical resolution.}
 \label{fig:pasta_fields}
\end{figure*}

The pair values are close: $S_{pn}=1.28\times10^{-4}$ in the slab and $1.13\times10^{-4}$ in the network. In both, the neutron density varies over the displayed proton surface (Fig.~\ref{fig:pasta_fields}). The slab is consequently not a null control. Although its connectivity remains that of a layer, its density develops an in-plane modulation during the HF iteration. The same transverse modes $(\pm2,\pm2)$ persist after refinement from $24^3$ to $30^3$; the maximum proton residual relative to the in-plane mean profile changes from $4.6\%$ to $5.0\%$. This persistence does not establish the physical origin of the modulation or its independence of cell size.

The factorization explains why the similar pair values do not imply similar triples. The weighted pair factors are $S_{pn}^{(W)}=1.27\times10^{-4}$ and $1.06\times10^{-4}$ for the slab and network, respectively. Their directional factors are $T_W=6.77\times10^{-7}$ and $2.98\times10^{-4}$, giving $S_{npW}=8.59\times10^{-11}$ and $3.16\times10^{-8}$. The larger separation is thus in the Coulomb contribution along the joint density levels, rather than in the weighted pair misalignment. The network triple is about 370 times the slab value on this mesh. This separation describes the computed configurations, while the mesh changes above limit its quantitative interpretation.

\subsection{Locating the pair signal by density level}

\begin{figure*}[t]
 \includegraphics[width=\textwidth]{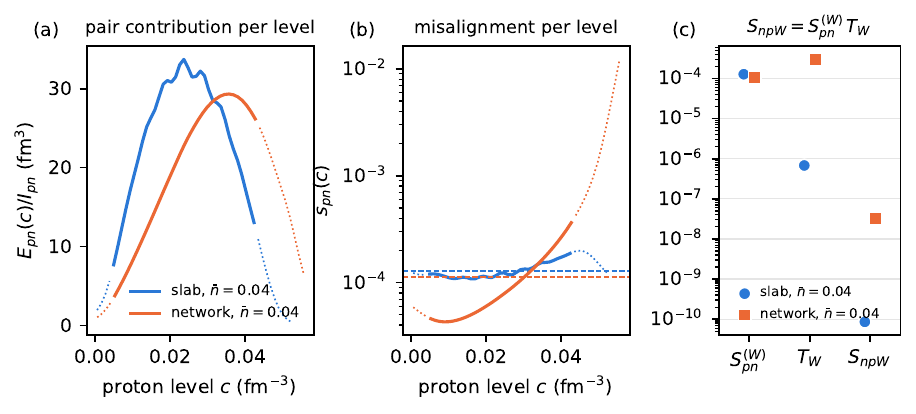}
 \caption{Preliminary level decomposition and triple factorization for the slab (circles) and network (squares) of Fig.~\ref{fig:pasta_fields}. (a) Pair contribution per proton level, $\mathcal E_{pn}(c)/I_{pn}$, evaluated over the whole periodic cell after Fourier interpolation to one quarter of the native spacing ($0.211\,\mathrm{fm}$), using 48 bins and Gaussian smoothing of width $\sigma=0.04 n_0Z/A$, with $n_0=0.16\,\mathrm{fm}^{-3}$. This is the same profile procedure used for the nuclei. (b) Level misalignment $s_{pn}(c)=\mathcal E_{pn}(c)/\mathcal D_{pn}(c)$; dashed lines give $S_{pn}$. Dotted portions mark the outer $2.5\%$ tails of the cumulative denominator weight. (c) The weighted pair factor, directional factor, and their product $S_{npW}=S_{pn}^{(W)}T_W$. Fourier interpolation improves the sampling of levels without adding physical resolution. The integrated values in (c) are evaluated on the native $30^3$ mesh.}
 \label{fig:pasta_levels}
\end{figure*}

The level decomposition distinguishes the two configurations despite their similar integrated pair values (Fig.~\ref{fig:pasta_levels}). The slab contribution extends over a broad range of proton levels, approximately $0.005$--$0.045\,\mathrm{fm}^{-3}$, while $s_{pn}(c)$ remains nearly constant and close to $S_{pn}$ over the levels with appreciable weight. In the network, the contribution is concentrated at higher proton levels, and $s_{pn}(c)$ rises from roughly $5\times10^{-5}$ to $4\times10^{-4}$. The slab thus has nearly level-independent misalignment; the network's misalignment depends more strongly on the density level.

Sampling the levels requires finer quadrature than the native solver grid. On that grid, the nearly one-dimensional slab places much of its pair signal in narrow bands set by the positions of the grid planes. Varying the histogram bins alone does not remove this sampling effect. We therefore evaluate the profiles on the Fourier-interpolated densities before binning and smoothing. Increasing this interpolation factor from four to six changes the profiles by about $1\%$ or less, and the profiles are stable over 32--96 bins. With the corrected estimator, the $24^3$ and $30^3$ normalized numerator profiles differ in $L^1$ norm by about $6\%$ for the slab and $7\%$ for the network. The interpolation checks establish the stability of the estimator on each representation; they do not establish convergence of the HF fields.

This preliminary comparison illustrates why the pair, its level profile, and the factorized triple should be read together: similar total misalignment can occupy different density levels and accompany a very different third-field contribution. It does not yet demonstrate an advantage over morphological descriptors for configurations of the same topology. That assessment requires further relaxation checks, spatial refinement, and cell-size tests.

\section{Scope of the diagnostic}
\label{sec:scope}

The tests above support a specific use of the invariants: comparing the directions in which fields vary and locating their disagreement. Their gradient weights emphasize regions where the fields vary together. They do not identify the microscopic cause of a difference.

A simpler diagnostic is the normalized correlation of the two gradients,
\begin{equation}
 C_{pn}=\frac{\int\nabla n_p\cdot\nabla n_n\,dV}{\bigl[\int|\nabla n_p|^2dV\int|\nabla n_n|^2dV\bigr]^{1/2}}
 =C_{\mathrm{mag}}\,\langle\cos\theta\rangle ,
 \label{eq:corr}
\end{equation}
with
\begin{gather}
 C_{\mathrm{mag}}=\frac{\int|\nabla n_p||\nabla n_n|\,dV}{\bigl[\int|\nabla n_p|^2dV\int|\nabla n_n|^2dV\bigr]^{1/2}},\\
 \langle\cos\theta\rangle=\frac{\int\nabla n_p\cdot\nabla n_n\,dV}{\int|\nabla n_p||\nabla n_n|\,dV}.
\end{gather}
The correlation equals one only when the gradients are related by a single positive constant throughout the domain. Its magnitude factor therefore responds to surface position and thickness even when the normals align and $S_{pn}=0$. In the static nuclei, this factor carries $\corrMagMin$--$\corrMagMax\%$ of $1-C_{pn}$ (Table~\ref{tab:corr}). The two measures can rank configurations differently: $1-C_{pn}$ places $^{76}$Ge below $^{128}$Xe, whereas $S_{pn}$ places it $\corrGeXe$ times higher.

In the magnetized cells, the magnitude factor carries at least $\corrCellMag\%$ of $1-C_{pn}$. For solutions with $|\Delta q_2^p|\leq\qTol$, $S_{pn}$ differs by up to a factor of $\qContrast$, while $1-C_{pn}$ changes by at most $\corrCellPairs\%$. Thus the correlation is dominated here by gradient magnitudes; the pair invariant isolates the angular disagreement, and its level decomposition locates it.

\begin{table}[t]
 \caption{Pair invariant and gradient correlation [Eq.~\eqref{eq:corr}] for the static nuclei at $1$ fm spacing: $1-C_{pn}$ and its magnitude ($1-C_{\mathrm{mag}}$) and angular ($1-\langle\cos\theta\rangle$) parts.}
 \label{tab:corr}
 \begin{ruledtabular}
 \begin{tabular}{lcccc}
 & $S_{pn}$ & $1-C_{pn}$ & $1-C_{\mathrm{mag}}$ & $1-\langle\cos\theta\rangle$\\
 $^{20}$Ne & $1.5\times10^{-5}$ & $1.5\times10^{-4}$ & $1.4\times10^{-4}$ & $9.8\times10^{-6}$ \\
$^{24}$Mg & $1.8\times10^{-5}$ & $2.1\times10^{-4}$ & $1.9\times10^{-4}$ & $1.6\times10^{-5}$ \\
$^{34}$Mg & $4.5\times10^{-3}$ & $3.7\times10^{-2}$ & $3.1\times10^{-2}$ & $6.3\times10^{-3}$ \\
$^{76}$Ge & $8.3\times10^{-4}$ & $9.7\times10^{-3}$ & $7.7\times10^{-3}$ & $2.0\times10^{-3}$ \\
$^{128}$Xe & $3.4\times10^{-4}$ & $1.1\times10^{-2}$ & $9.5\times10^{-3}$ & $1.9\times10^{-3}$ \\

 \end{tabular}
 \end{ruledtabular}
\end{table}

A stronger test asks whether the pair invariant distinguishes nearly degenerate states that also have similar low moments. The two unconstrained $B=0$ minima differ in $S_{pn}$ by a factor of $\nearBzero$, but already have quadrupole moments of opposite sign. Among the $\nearPairs$ same-field pairs within $10$ keV per nucleon and with $|\Delta q_2^p|,|\Delta q_4^p|\leq0.01$, the largest $S_{pn}$ ratio is $\nearMax$. This is within the starting-basis spread (up to $\nearSpread\%$) and mesh dependence ($\chkMeshSeven\%$). The cells therefore provide no resolved distinction between near-degenerate states beyond their low moments.

The invariants require no origin or axis; the independent surface-radius and axial-departure checks used here do. Multipoles describe each field about a center, Minkowski functionals and tensors describe individual surfaces~\cite{Schuetrumpf2013,Schneider2013,SchroederTurk2013}, and joint contour nets describe the topology of joint levels~\cite{Duke2012}. The present comparison measures how fields vary across those levels. The periodic comparison in Sec.~\ref{sec:pasta} illustrates this distinction for a slab and a connected network. Its spatial and cell-size convergence, and its advantage over morphological descriptors for configurations of the same topology, remain unestablished.

\section{Conclusions}
\label{sec:conclusions}

The pair invariant measures how one field varies over the level surfaces of another. In the magnetized cells it distinguishes solutions with comparable proton quadrupole moments; in the static nuclei its level decomposition locates the larger $^{34}$Mg signal across the proton surfaces. The pair contrast persists under the grid and box changes tested. A complete multipole expansion can recover the same information, but the invariant provides the comparison directly from the fields.

The triple adds variation along the intersections of two level surfaces. Its factorization is essential to interpretation: a small triple may reflect either weak pair misalignment or little variation of the third field along those curves. The directional factor separates the sampled triaxial and prolate configurations. In the collision it accounts for the rise before density contact, also reproduced by rigidly approaching static fragments, while the weighted pair term remains nearly constant.

These results establish a way to measure and locate relative field geometry without choosing an origin or axis. They do not establish a measure of skin thickness or triaxiality. A resolved nonzero triple rules out a common axis; zero remains inconclusive. The cells show no resolved distinction between near-degenerate states beyond that supplied by their low moments. Collision convergence beyond $110$ fm/$c$ remains to be tested. The preliminary periodic-pasta comparison shows similar density-pair signals with different level distributions and directional factors in a slab and a network. These observations concern the stored $30^3$ fields; further relaxation, spatial refinement, and cell-size tests are required before drawing quantitative conclusions about pasta.

\section*{Data availability}
The data that support the findings of this article, the analysis code with its symbolic and numerical tests, and the scripts that produce the figures and tables are openly available in Zenodo~\cite{LimaZenodo2026}. Each derived data file records the SHA-256 checksums of the solutions it was derived from. The relativistic mean-field solver is not part of the archive; the fields of all its solutions are included. Sky3D is publicly available~\cite{Maruhn2014}.

\begin{acknowledgments}
R.C.R.L. thanks CENAPAD-SP and SciCom-UFES for
the computational allocation.
\end{acknowledgments}

\appendix
\section{Curvilinear evaluation and analytic checks}
\label{app:axial}

\subsection{Derivatives and quadrature}

The axial implementation stores the fields at the centers of finite-volume cells of a mesh in orthogonal coordinates $(u,\theta)$ of the meridian plane, uniform in $u$ and in $\theta$, with $\eta=\cos\theta$. In the spherical cell $u$ and $\theta$ coincide with the radius $r$ and the polar angle to within $5\times10^{-4}$ at the innermost nodes ($r\leq0.2$ fm) and to better than $10^{-6}$ beyond $2$ fm. The mesh covers the half-cell $z\geq0$; the symmetry axis and the reflection plane are cell faces, so no node lies on them.

Derivatives with respect to $u$ and $\theta$ are second-order centered differences on the uniform mesh,
\begin{equation}
 \partial_uf\big|_{i,j}\simeq\frac{f_{i+1,j}-f_{i-1,j}}{2\Delta u},
\end{equation}
and analogously in $\theta$, with second-order one-sided differences at the first and last nodes of each line. The physical components of the gradient are $(\partial_uf/h_u,\ \partial_\theta f/h_\theta)$, with the scale factors of the mesh coordinates evaluated analytically at the nodes; they reduce to the spherical values $h_r=1$ and $h_\theta=r$ within the accuracy quoted above. Fields that are smooth functions of $\varrho^2$ and $z$ are smooth in $(u,\theta)$, so the differences converge at second order. The quadrature weights are the exact volumes of the finite-volume cells, doubled to account for $z<0$. With mesh spacing $h$ the error of $I_{ab}$ behaves as $Ch^2$, and from two meshes with spacings $h$ and $2h$ we form the Richardson estimate
\begin{equation}
 I_{ab}^{(R)}=\frac{4I_{ab}(h)-I_{ab}(2h)}{3}.
\end{equation}

\subsection{Analytic checks}

\begin{figure}[tbp]
 \includegraphics[width=\columnwidth]{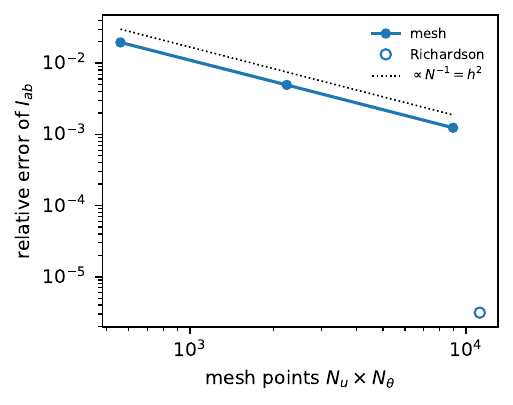}
 \caption{Relative error of $I_{ab}$ for an analytic pair of fields against the number of points of the spherical-cell mesh. Open symbol: Richardson extrapolation from the two finest meshes. Dotted line: second-order convergence.}
 \label{fig:conv}
\end{figure}

The identity $|\partial(\varrho,z)/\partial(u,v)|=h_uh_v$ and the orthogonality of the metric, on which Eq.~\eqref{eq:jac} relies, were verified symbolically for cylindrical and spherical polar coordinates (\symOK{} checks pass). Numerical tests with analytic fields on the solver mesh of the spherical cell give, on $160\times56$ points, $S=\valSrad$ for two radial fields, $S=\valSdep$ for a field and its cube, and $S=1$ to seven digits for gradients that are orthogonal everywhere ($f_a=\varrho^2$, $f_b=z$). For an anisotropic Gaussian pair with known $I_{ab}$, the relative error is $\valErrA$, $\valErrB$, and $\valErrC$ on meshes of $40\times14$, $80\times28$, and $160\times56$ points; it decreases at second order (ratio of the last two errors $\valRatio$; Fig.~\ref{fig:conv}), and Richardson extrapolation from the two finest meshes reproduces the exact value to $\valRich$. For a spherical Thomas--Fermi solution of the full mean-field problem, all invariants vanish to $\TFnull$.

\section{Dirac solutions in the magnetized cell}
\label{app:dirac}

The solutions of Sec.~\ref{sec:cells} describe neutral neutron--proton--electron matter at mean baryon density $\bar n_B=0.019\,\mathrm{fm}^{-3}$ and proton fraction $Y_p=0.3$ in a spherical Wigner--Seitz cell of radius $R=16.73$ fm, with the NL3 parameterization~\cite{Lalazissis1997}. The scalar, vector-isoscalar, and vector-isovector meson fields and the Coulomb potential satisfy the static field equations with sources given by the local densities, on the $\solMesh$ mesh of Appendix~\ref{app:axial} with Neumann conditions at the cell surface. Neutrons and electrons are treated in the relativistic Thomas--Fermi approximation, electrons in Landau levels of the uniform field $\bm B=B\hat z$; their chemical potentials fix the mean densities $(1-Y_p)\bar n_B$ and $Y_p\bar n_B$, without beta equilibrium. Anomalous magnetic moments, the Dirac sea, pairing, and the spatial components of the vector-meson fields (and hence their feedback from the proton currents) are not included.

The protons obey the Dirac equation with minimal coupling,
\begin{equation}
 \bigl[\bm\alpha\cdot(\bm p-e\bm A)+\beta M^*(\bm r)+V_p(\bm r)\bigr]\psi_i=\varepsilon_i\psi_i,
\end{equation}
with $\bm A=\tfrac{B}{2}(-y,x,0)$, $M^*=M-g_\sigma\sigma$, and $V_p$ the vector-meson and Coulomb potentials. Axial and reflection symmetry make $\Omega=j_z$ and parity good quantum numbers. The large components are expanded in eigenfunctions of an oscillator confined to the cell, $-\nabla^2+\varrho^2/b_\perp^4+z^2/b_z^4$ with Dirichlet conditions at $r=R$, obtained from a spherical Bessel basis~\cite{Than2011} with $k_{\max}=\solKmax\,\mathrm{fm}^{-1}$. The eigenfunctions are kept up to the energy $(2N_F+3)/b_0^2$ with $N_F=\solNF$, the cut that a free oscillator basis of $N_F$ shells would have, with $b_0^3=b_\perp^2b_z$; the confined basis is not literally a set of oscillator shells. The length $b_0=(M^*\hbar\omega)^{-1/2}$ uses $\hbar\omega=41A^{-1/3}$ MeV, the aspect ratio $b_z/b_\perp$ is that of the starting basis, and $b_\perp$ includes the magnetic confinement, $b_\perp^{-4}\to b_\perp^{-4}+(eB/2)^2$. The small components are built by restricted kinetic balance, $\sigma\cdot\bm\pi$ applied to the large-component functions; they do not vanish at the cell surface in general. Only eigenvalues between $0$ and $M+10$ MeV are retained. Levels are occupied with a Fermi--Dirac distribution of width $kT=\solKT$ MeV, with the proton number fixed by the chemical potential. The proton densities enter the field equations, and the procedure is iterated, with linear mixing $\solMix$, until the residuals of the densities and fields fall below $10^{-7}$. Deformed configurations are obtained by minimizing $E+\lambda\langle\hat Q\rangle$ with $\hat Q=(2z^2-\varrho^2)f(r)$, $f(r)=\{1+\exp[(r-0.75R)/(1\,\mathrm{fm})]\}^{-1}$, for $\lambda$ between $-0.02$ and $0.02$ MeV fm$^{-2}$. Each calculation starts once from a prolate and once from an oblate basis, at $B=0$, $1$, $2$, $3$, $5$, $7$, and $10\times10^{17}$ G.

Three recomputed solutions test the basis and the mesh, against the unconstrained solutions from the prolate starting basis. At $7\times10^{17}$ G this is the selected lower-energy solution; at $10^{18}$ G the selected one comes from the oblate start, whose $S_{pn}$ differs by $\startTen\%$, and the checks do not test it directly. With $N_F=24$ shells, $S_{pn}$ changes by $\chkBasisSeven\%$ at $7\times10^{17}$ G and $\chkBasisTen\%$ at $10^{18}$ G, and the other eight pairs of Fig.~\ref{fig:pairs} by at most $\chkPairsBasis\%$. On a $120\times40$ mesh at $7\times10^{17}$ G, $S_{pn}$ changes by $\chkMeshSeven\%$ and the other pairs by at most $\chkPairsMesh\%$. The energy per nucleon changes by at most $\chkEA$ keV and the proton quadrupole moment by at most $\chkQp\%$.

The starting basis matters more at weak fields. At $B=0$ the two unconstrained calculations converge to distinct minima, oblate ($q_2^p=\qZeroOb$) and prolate ($q_2^p=\qZeroPr$, lower by $\dEZero$ keV per nucleon), with $S_{pn}=\startZeroOb$ and $\startZeroPr$. At $1$ and $2\times10^{17}$ G both reach the same shape, but $S_{pn}$ differs by $\startLow$--$\startLowMax\%$; from $3\times10^{17}$ G on, by at most $\startHigh\%$. Figure~\ref{fig:pairs} and the values quoted for individual fields use the unconstrained solution of lower energy; Fig.~\ref{fig:beyond} shows all solutions.

\bibliography{references}
\end{document}